\pdfoutput=1

\documentclass[11pt]{article}

\usepackage[preprint]{acl}

\usepackage{times}
\usepackage{latexsym}

\usepackage[T1]{fontenc}

\usepackage[utf8]{inputenc}

\usepackage{microtype}

\usepackage{inconsolata}

\usepackage{graphicx}
\usepackage[most]{tcolorbox}
\tcbset{
    frame code={},
    center title,
    left=0pt,
    right=0pt,
    top=0pt,
    bottom=0pt,
    colback=gray!12,
    colframe=white,
    width=\dimexpr\columnwidth\relax,
    enlarge left by=0mm,
    boxsep=5pt,
    arc=0pt, outer arc=0pt,
    }

\title{Creating a Taxonomy for Retrieval Augmented Generation Applications}

\author{
 \textbf{Irina Nikishina\textsuperscript{1}},
 \textbf{Özge Sevgili\textsuperscript{1,2}},
 \textbf{Mahei Manhai Li\textsuperscript{3}},
 \textbf{Chris Biemann\textsuperscript{1,2}},
 \textbf{Martin Semmann\textsuperscript{2}},  \\
\\
\\
 \textsuperscript{1}Language Technology Group, University of Hamburg, Germany \\
 \textsuperscript{2} HCDS Group, University of Hamburg, Germany \\
 \textsuperscript{3} Information Systems, University of Kassel, Kassel, Germany
\\
 \small{
   \textbf{Correspondence:} \href{mailto:irina.nikishina@uni-hamburg.de}{irina.nikishina@uni-hamburg.de}
 }
}

\begin{document}
\maketitle
\begin{abstract}
In this research, we develop a taxonomy to conceptualize a comprehensive overview of the constituting characteristics that define retrieval augmented generation (RAG) applications, facilitating the adoption of this technology for different application domains. To the best of our knowledge, no holistic RAG application taxonomies have been developed so far. We employ the method foreign to ACL and thus contribute to the set of methods in the taxonomy creation. It comprises four iterative phases designed to refine and enhance our understanding and presentation of RAG's core dimensions. We have developed a total of five meta-dimensions and sixteen dimensions to comprehensively capture the concept of RAG applications. Thus, the taxonomy can be used to better understand RAG applications and to derive design knowledge for future solutions in specific application domains. 
\end{abstract}

\section{Introduction} \label{sec:introduction}
Large Language Models (LLMs) have been identified to have several core limitations. These include a tendency to generate incorrect or misleading information (hallucinations) \cite{liang-etal-2024-learning,nonkes-etal-2024-leveraging}, poor arithmetic capabilities, a lack of interpretative power, the high costs associated with model revisions, limitations in handling less popular or low-resource concepts and entities, and an inability to reference sources accurately \cite{Barnett2024Seven,Soudani2024Fine,Zhao2024Retrieval-Augmented}. Several approaches have been developed to mitigate the limitations of LLMs, while retrieval augmented generation (RAG) is as of now deemed as one of the most promising \cite{Gao2024Retrieval-Augmented}. RAG primarily enhances LLMs by incorporating contextual information during the retrieval process, significantly improving the generated content's accuracy and consistency. Consequently, RAG improves LLM tasks and applications in various ways, as evidenced by recent studies \cite{Asai2023Self-RAG:,Jiang2023Active,Martino2023Knowledge}. Given its potential, this paper aims to develop a conceptualization for RAG applications, illustrating how RAG can be systematically implemented to improve LLM tasks and applications across various domains. Recent studies, such as those by \citet{Asai2023Self-RAG:,Jiang2023Active,Martino2023Knowledge}, have shown various ways in which RAG can enhance LLMs, underscoring its attributes as explainable, scalable, and adaptable in nature \cite{Siriwardhana2023Improving}. 


Given this value of RAG for real-world applications, still there is a dearth of systematization of the field. This is particularly evident in surveys, which emphasize technological aspects over practical applications \cite{Zhao2024Retrieval-Augmented}. Therefore, we aim to create a taxonomy that conceptualizes a comprehensive overview of the constituting characteristics that define RAG applications, facilitating the adoption of this technology. 
Current research on RAG is distributed across various disciplines, and since the technology is evolving very quickly, its unit of analysis is mostly on technological innovations, rather than applications in business contexts. 
To the best of our knowledge, there have not been any holistic RAG application taxonomies. Thus, our research question is as follows: \textit{``How can RAG applications be conceptualized in a taxonomy?''}

Therefore, the main contributions of the paper are as follows:

\begin{itemize}
    \item We present a RAG taxonomy offering a comprehensive framework to define and categorize the core characteristics of Retrieval-Augmented Generation (RAG) applications, promoting their broader adoption and practical use.
    \item We contribute to taxonomy creation methods within the ACL community by adapting the approach of \citet{Nickerson2013method} from the field of Information Systems.
    \item We validate the practical applicability of the created taxonomy by  reviewing various papers on RAG applications and also by using the automatic domain classification with ChatGPT.
\end{itemize}

\section{Related Work}

In this section, we discuss the peculiarities of the current survey before the other survey papers \cite{Zhao2024Retrieval-Augmented,Gao2024Retrieval-Augmented,
Li2022Survey,zhao-etal-2023-retrieving} and discuss the origins of our taxonomy creation approach.

While there exist already various surveys on RAG, we would like to specify, how the current survey paper is different from others. 
\citet{Li2022Survey} provide one of the first survey papers on the topic that covers early works on RAG with application to NLP tasks such as Dialogue Systems, Neural Machine Translation, Paraphrase Generation, Text Style Transfer, etc. In \citet{Zhao2024Retrieval-Augmented}, the authors comprehensively review existing efforts that integrate RAG techniques into AI-generated content scenarios, exhibiting how RAG contributes to current generative models. \citet{Gao2024Retrieval-Augmented} contextualize the broader scope of RAG research within the landscape of LLMs. The paper presents various RAG components, datasets, and evaluation setups being the perfect source for both understanding the RAG concept as well as delving into the topic. \citet{zhao-etal-2023-retrieving} specialize in multimodal research for images, videos, code, text, etc. The main idea is to structure our knowledge of RAG by creating a taxonomy for different characteristics of the RAG system in order to make their development easier for further RAG applications.

Classification research and typologies are used for the scientific pursuit of differences to theorize about commonalities \cite{Beaulieu2015Conceptual}. Classification schemes and theories of typologies originate in biology to study and classify species, but have since gained widespread adoption in application-oriented research domains like the information systems community \cite{Nickerson2013method}. While sometimes typologies are synonymous with the term framework, in this paper we will use the term taxonomy to structure the novel technological artifact known as RAG. Thus, we describe our methodology to develop a RAG application taxonomy following a systematic approach based on  \citet{Nickerson2013method}.

\section{Taxonomy Development} \label{sec:taxonomy}
In this section, we describe the methodology used for developing the taxonomy of RAG. We describe the process of paper selection, as well as the iterations of the methodology applied. Moreover, we describe the additional approach for identifying the domains where RAG was applied – an analysis of papers found with specific queries in Google Scholar and ACL Anthology using ChatGPT. 

\subsection{Methodology} \label{subsec:taxonomy-methodology}

Following the approach of \citet{Nickerson2013method}, we defined our meta-characteristic as ``structure and applications of retrieval augmented generation''. In doing so, we included conceptual work and case studies that aimed at specific application domains or use cases. Due to the high dynamic in the research area, we also included pre-print articles. To ensure minimal quality standards, we reviewed each article by two independent researchers. To define a level of saturation, we used the objective and subjective ending conditions as proposed by \citet{Nickerson2013method}. So, we examined a representative sample of the most recent literature, and those dimensions were stable for an iteration. Thus, no extensions, merges, or splits of characteristics were performed. We also ensured that dimensions have at least one characteristic, and those are directly derived from papers while being unique. Subjective ending conditions were considered as proposed. We aimed for comprehensiveness, robustness, conciseness, extensibility, and explainability. This is reflected in the adaptation of dimensions throughout the iterations as we joined, reorganized, and split categories. As we tackle a recent and ever-changing phenomenon, we conclude that an expanded set of dimensions, namely 16, is still useful for research and practice in the current state. Regarding robustness, we checked for a strict separation between dimensions as well as characteristics. Comprehensibility is ensured by our extensive approach. Extensibility and explainability were tackled by repeatedly applying examples to the taxonomy. 

To start the development process, we choose a twofold approach. First, we used aspects of conceptual-to-empirical to catalyze the initial iteration \cite{Nickerson2013method}. This was done by identifying relevant domains with the help of ChatGPT. Second, we also incorporated an empirical-to-conceptual approach that specifically used surveys of RAG, as those already provide some cumulative knowledge of the field. Still, most are rather new and have not gone through a proper double-blind review process. Thus, in combining both approaches, we propose a new angle to deal with emerging topics. Afterward, we strictly follow the empirical-to-conceptual regime.

\subsection{Paper Selection} \label{subsec:taxonomy-paper-selection}
As the object of interest is relatively novel, we employed a naïve approach for identifying and selecting papers. As a first step, we started with the search string ``RAG \& application'' 
at Google Scholar. The results encompassed several articles from the Association of Computational Linguistics, which is a driver of the general development of LLMs, as well as RAG. Thus, we deemed the general search strategy as useful. Despite high-quality results, most papers are in pre-print status and have not been part of a thorough peer review. Still, we include those, after quality checks by the author team. If articles are perceived as questionable by one of the researchers, we employed a cross-check by at least another researcher on the team. Also, we excluded published work with publishers that do not comply with the minimal standards of our research community. 

Due to the iterative approach taken, we reviewed twenty-eight papers in the taxonomy development process. While 20 papers have been reviewed and published in several venues, 8 papers are still in pre-print phase. However, we need to emphasize that all 8 papers were written in 2024 which means that they still can be accepted in the future.

\subsection{ChatGPT Domain Identification} \label{subsec:taxonomy-chatgpt}
To facilitate the time-consuming process of paper analysis, we decided to apply it for identifying the application domains and further compare the outcome to see, whether such an automated technique applies to the taxonomy construction. 
First, we created two queries for Google Scholar – a large search system for academic papers\footnote{All searches are done on a date, 11.09.2024}. The first query ``application of rag'' for the papers dating from 2023 (anywhere in the article) returned ninety-three results. The ``rag application'' query from 2023 (anywhere in the article) returned seventy-four papers. Additionally, we searched these two queries in ACL Anthology\footnote{\url{https://aclanthology.org/}}, and the first one returned two results, while \ five results are returned with the second one.
Then, inspired by \citet{Rafailov2023Direct}, we created a prompt to ChatGPT to cluster the extracted papers into domains. We formulated the prompt as follows: 


\begin{tcolorbox}
\begin{verbatim}
You are a scientific assistant writing 
a survey. Here below is a list of paper 
names. Your task is to cluster those pa-
pers into domains. Name those domains 
(it might be something like NLP, 
or medicine).   
\end{verbatim}
\end{tcolorbox}

After the prompt, we pasted 
names from the first query separated with the line separator. Based on the titles provided, the ChatGPT model identified the following 10 classes: (1) Artificial Intelligence and Natural Language Processing (AI \& NLP);
(2) Cybersecurity; (3) Medicine and Healthcare;
(4) Business and Economics;
(5) Education and Programming; (6) Social Sciences and Ethics; (7) Disaster and Risk Management; (8) Physics and Engineering; (9) Data Science and Knowledge Discovery; (10)
Adversarial AI and Machine Learning.

When providing ChatGPT papers from the ``rag application'' query, the output is: (1) Artificial Intelligence and Language Models; (2) Legal and Justice; (3) Medicine and Healthcare; (4) Education and Pedagogy; (5) Engineering and Construction; (6) Data Science and Information Retrieval.

It is also important to emphasize that in addition to the class names, the model returned paper examples for each class, therefore, we were able to primarily check the correctness of the identified classes. Furthermore, during iterations, we expected to use a manual paper check to prove the ChatGPT clustering efficiency. 

\subsection{Iterations} \label{subsec:taxonomy-iterations}
We performed four iterations to build our initial RAG application taxonomy. Overall, we analyzed $28$ papers, including 4 surveys on RAG, that already performed the extensive analysis and generalization of previous works \cite{Li2022Survey, zhao-etal-2023-retrieving, Gao2024Retrieval-Augmented, Zhao2024Retrieval-Augmented}. Those papers already comprise over 2000 
citations, resulting in more than twenty-eight papers involved in our study in total. Moreover, a major part of the dimensions was added during the first iteration, where two survey papers were analyzed. Therefore, we consider this number of iterations reasonable, which was also confirmed by meeting the following objective condition – no new dimensions or characteristics were added, merged, or split in the iteration.  

\begin{figure*}[ht!]
\centering
\vspace{0.5cm}
\includegraphics[ width=\textwidth]{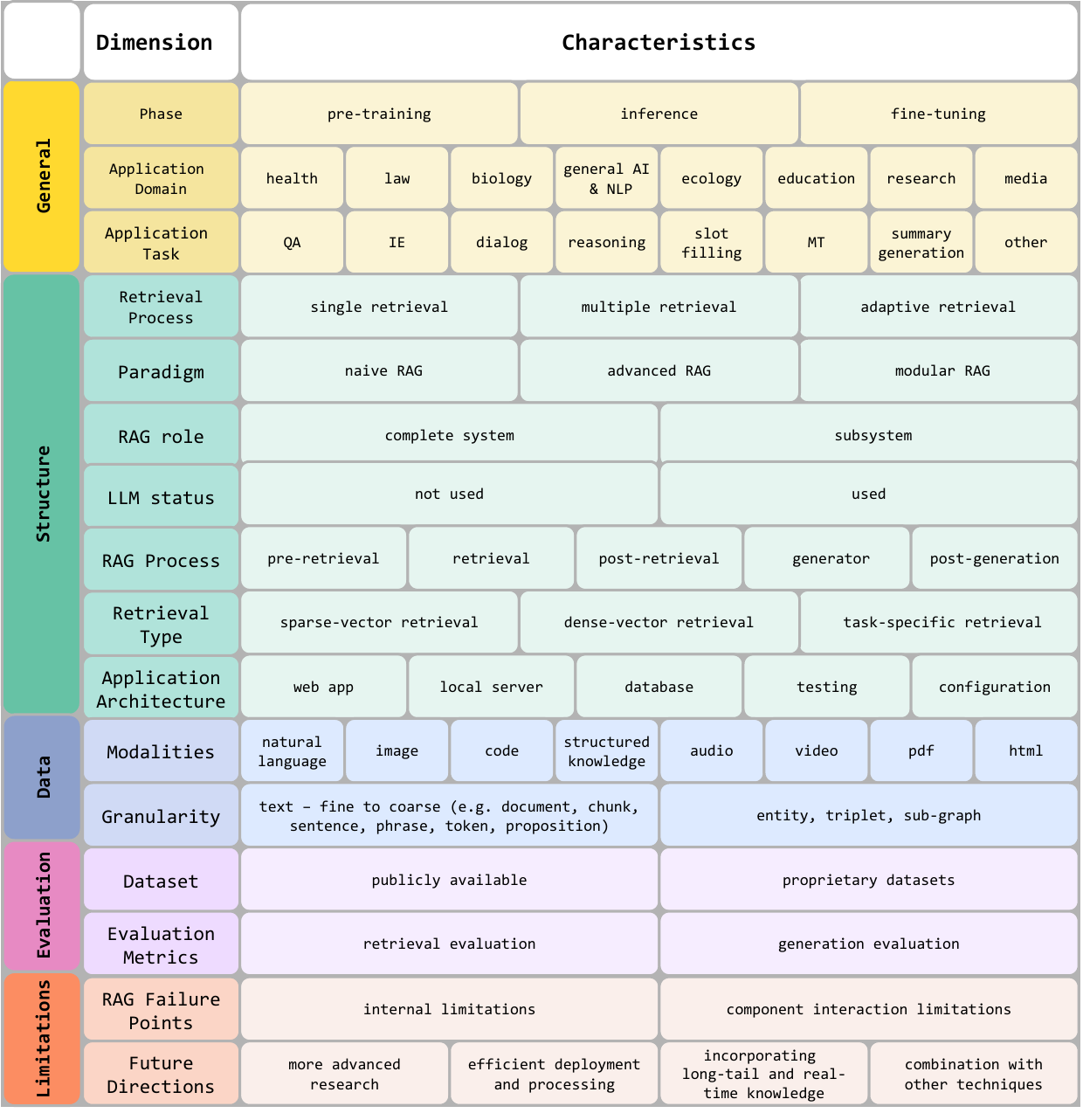}
\caption{RAG Taxonomy created from twenty-eight papers within four iterations.}
    \label{fig:taxonomy}
\end{figure*}

Our iterative development process for the RAG application taxonomy, as illustrated by Figure \ref{fig:iterations} in Appendix \ref{sec:appendix} started with an initial set of eleven dimensions, mapped across five meta-dimensions. In this first iteration, the key dimensions of the RAG Phase, Application Domain, Application Task, RAG Process, Paradigm, Retrieval Type, Retrieval Process, RAG Role, Modality, Evaluation Metrics, and Failures of RAG were established, each annotated with a specific number of characteristics indicated by the numbers in parentheses. In the second iteration, we enriched the taxonomy by analyzing seven additional papers, leading to the introduction of 
three new dimensions: LLM Status, Granularity, and Dataset
. Concurrently, we refined several existing dimensions by either merging or adding new characteristics, reflecting deeper insights and broader coverage. By the third iteration, only two new dimensions were necessary: Application Architecture and Future Directions, indicating a nearing saturation in the scope of the taxonomy. Adjustments were made to five dimensions during this phase, demonstrating a trend toward the stabilization of the taxonomy's structure. The fourth iteration confirmed the saturation, as no new changes were made to the taxonomy, suggesting that the existing structure sufficiently captured the relevant aspects of RAG applications as evidenced by the literature. Across all iterations, the taxonomy has evolved to accommodate and anticipate the dynamic nature of RAG applications, ensuring its relevance and utility in future research. 

\section{Results} \label{sec:results}
In this section, we present the final RAG taxonomy created from twenty-eight papers in four iterations.  Figure \ref{fig:taxonomy} illustrates the entire taxonomy divided into five main components. In the following subsections, we describe each component in more detail.  

\subsection{General} \label{subsec:results-general}
The first group of dimensions is devoted to general aspects of RAG systems, regardless of specific structural aspects. Within this group, we identified three dimensions to represent the general aim and motivation of applied RAG. 

\paragraph{$\bf D_1$ Phase:} The dimension subsumes the focus of RAG in place. It also relates to the evolving discourse – despite being a novel phenomenon. Research shows that there are three primary areas of application for RAG. The resulting characteristics are pre-training, inference, and fine-tuning \cite{Gao2024Retrieval-Augmented}.

\paragraph{$\bf D_2$ Application Domain:} In total, eight application domains are found in the discourse. The most frequent are health, law, biology, general AI and NLP, as well as ecology \cite{Gao2024Retrieval-Augmented}. Further application domains are education and research \cite{Barnett2024Seven} as well as media \cite{Siriwardhana2023Improving}. 

\paragraph{$\bf D_3$ Application Task:} RAG methods can be applied to different applications or downstream tasks. In this dimension, the characteristics are the prominent tasks \cite{Gao2024Retrieval-Augmented}. We consider eight characteristics, i.e. Question-Answering (QA), Information Extraction (IE), Dialog, Reasoning, Slot Filling \cite{Glass2021Robust}, Machine Translation \cite{Li2022Survey}, Summarization \cite{Zhao2024Retrieval-Augmented}, others. QA has several different types, e.g. open domain QA, abstractive QA \cite{Lewis2021Retrieval-Augmented}, GraphQA \cite{He2024G-Retriever:}, etc. Other contains some other tasks, for example, Fact Checking/Verification, Question Generation \cite{Lewis2021Retrieval-Augmented}, Code search \cite{Gao2024Retrieval-Augmented}, and many more. 
\subsection{Structure} \label{subsec:results-structure}
This includes an examination of the underlying technologies that form the architecture of the RAG application, determining whether the RAG acts as the principal system or merely a component within a larger system. We further delineate the structure of RAG systems by analyzing different RAG paradigms—such as naive, advanced, and modular RAG—which reflect varying levels of complexity and integration. Additionally, we consider the specific contexts or processes where the RAG retrieval is realized. In total, the structure includes key characteristics that distinguish RAG systems.  

\paragraph{$\bf D_4$ Retrieval Process:} The retrieval process represents to what extent the RAG uses retrieval. Single retrieval, multiple retrieval, and adaptive retrieval are the identified characteristics \cite{Gao2024Retrieval-Augmented}. Single retrieval thus solely relies on a single retrieval sequence in a RAG, while multiple retrieval is an iterative or sequential approach. Adaptive retrieval is the most contextual approach, as it integrates the results of prior retrieval to adapt the next iteration of retrieval. 

\paragraph{$\bf D_5$ Paradigm:} \citet{Gao2024Retrieval-Augmented} categorize the RAG research paradigm into three Naive RAG, Advanced RAG, and Modular RAG. In this dimension, these three categories are the characteristics. In Naive RAG, there are three parts, i.e., indexing, retrieval, and generation. Advance RAG also includes pre-retrieval and post-retrieval parts before and after retrieval. Modular RAG provides flexibility with different modules, e.g., search module, memory module, etc. 

\paragraph{$\bf D_6$ RAG Role:} Within the application landscape, the role of RAG systems can vary significantly: they can operate as dedicated, monolithic systems or as modular components integrated within other application systems \cite{Zhao2024Retrieval-Augmented}. According to the authors, subsystems can be part of larger architectures that employ multiple frameworks, i.e. RetDream for 3D Generation \cite{Seo2024Retrieval-Augmented}, R-ConvED for video captioning \cite{Chen2023Retrieval}, and kNN-TRANX \cite{Zhang2023Syntax-Aware} for text-to-code tasks. In the above-mentioned systems, RAG is used as an additional step to the pipeline, enhancing generation with the retrieved data. 
\paragraph{$\bf D_7$ LLM Status:} This dimension is binary, it checks for the adaptability of the LLM in place. So it can either be not used, meaning no further approach is taken to improve the LLM performance, or it can be used \cite{Chen2024Benchmarking}. Used thus leads to different forms. It can be trainable to be adjusted in each context or it can be looped as a specification of the paradigm modular RAG. 

\paragraph{$\bf D_8$ RAG Process:} In this dimension, processes in RAG models are discussed mostly based on the information by \citet{Gao2024Retrieval-Augmented}. Five characteristics are considered, i.e., pre-retrieval, retrieval, post-retrieval, generation, and post-generation. The pre-retrieval step involves some techniques applied before the retrieval step, for instance, chunking, vectorizing, indexing, and some other strategies to e.g., optimize indexing, enhance user input, etc. In the retrieval step, the relevant information to the user input is retrieved. Post-retrieval includes methods to improve the retrieved information during integration with user input, e.g. re-rank the information or subgraph construction \cite{He2024G-Retriever:}. In the generation step, LLM provides a response to the prompt that contains the retrieved information and user input. Post-generation contains strategies that can be applied after generation, e.g. output rewrite \cite{Zhao2024Retrieval-Augmented}. Note that there exist various modules to enhance different components (see \citet{Gao2024Retrieval-Augmented} for more information). 

\paragraph{$\bf D_9$ Retrieval Type:} There are different types of retrieval augmentation methods \cite{Li2022Survey}. In this dimension, three characteristics are considered, i.e., sparse-vector retrieval, dense-vector retrieval, and task-specific retrieval \cite{Li2022Survey}. Sparse-vector retrieval involves methods, e.g., TF-IDF, BM25, etc. Dense-vector retrieval contains models based mostly on low-dimensional dense vectors, e.g., BERT-encoders, and relying often on vector databases. In task-specific retrieval, the retrieval module is based on task \cite{Li2022Survey} and might comprise a database \cite{Radeva2024Web}. Some research works directly using the edit distance between natural language texts \cite{Hayati2018Retrieval-Based} or abstract syntax trees (AST) of code snippets \cite{Poesia2021Synchromesh:}. 

\paragraph{$\bf D_{10}$ Application Architecture:} When developing a RAG system, in addition to the RAG structure, we also need to consider the structure of the final application and the interaction of the components. \cite{Radeva2024Web} present a web application RAG system that consists of the ``local or server-based installation'', ``web application'', ``vector storage'' (database), as well as the testing and configurations. Therefore, this dimension comprises the web app, local server, database, testing, and configuration characteristics. 

\subsection{Data} \label{subsec:results-data}
\paragraph{$\bf D_{11}$ Modalities:} Although the concept of RAG was originally developed for text-based generation, its use has been adapted for a variety of other generation modalities 
\cite{Chen2024Benchmarking,Gao2024Retrieval-Augmented,Lewis2021Retrieval-Augmented,Zhao2024Retrieval-Augmented}. This includes programming code, audio, visual content, such as images and videos, 3D models, and other knowledge structures. The latter can include table structures, higher-level modeling languages, graphs, textual graphs \cite{He2024G-Retriever:}, or knowledge graphs. The fundamental principles of RAG remain similar across these different modalities, even though slight modifications to the augmentation methods are sometimes required.  

\paragraph{$\bf D_{12}$ Granularity:} This dimension is for different granularities of retrieved data based on the information by \citet{Gao2024Retrieval-Augmented}. The modality can be natural language (or text), yet still, the retrieved granularity might vary from fine to coarse, e.g., document, chunk, sentence, proposition, etc. \cite{Gao2024Retrieval-Augmented}. Similarly, there exist several granularities in structured data, e.g., sub-graph, triplet, entity, etc. \cite{Gao2024Retrieval-Augmented}. 

\subsection{Evaluation} \label{subsec:results-evaluation}
\paragraph{$\bf D_{13}$ Dataset:} Regarding the datasets used for RAG, most RAG surveys consistently list the datasets used regardless of the application task, the RAG step to be evaluated on as well as the dataset availability. Thus, we focused on the matter of availability and considered two characteristics, i.e., publicly available, and proprietary datasets. Some examples of publicly available datasets are e.g. FEVER \cite{Thorne2018FEVER:}, SQuAD \cite{Rajpurkar2016SQuAD:} etc., and the dataset, e.g. by \citet{Bondarenko2020Comparative}, is an example for proprietary datasets. 

\paragraph{$\bf D_{14}$ Evaluation Metrics:} When reviewing papers discussing separate models and architectures, we can see that the authors mostly use task-specific metrics \cite{Thakur2024Loops} or the generation output quality only \cite{Chen2024Benchmarking}. However, \citet{Gao2024Retrieval-Augmented} split evaluation metrics into two groups: retrieval evaluation and generation evaluation metrics, which are the base parts of RAG. The first group evaluates the relevance of the retrieved data to the query and is mostly represented with the ranking evaluation metrics: Precision@k, Recall@k, F@1, MRR, MAP \cite{Gao2024Retrieval-Augmented}. The second group involves generation evaluation metrics, such as BLEU, METEOR, ROUGE, PPL \cite{Radeva2024Web} and Accuracy, Rejection Rate, Error Detection Rate, Error Correction Rate \cite{Chen2024Benchmarking}. 

\subsection{Limitation} \label{subsec:results-limitation}
Despite multiple advantages and ubiquitous application, \citet{Zhao2024Retrieval-Augmented} outline limitations and possible directions of RAG. We describe the last two dimensions in more detail, also considering failures from \citet{Barnett2024Seven}. 
\paragraph{$\bf D_{15}$ RAG Failure Points:} RAG limitations can be divided into two groups: internal (related to the system component efficiency) and integration (related to the problems of RAG components interaction). Here, we discuss each type separately.  

The most evident and the most frequent failure point for RAG is the retrieval step. Noises in retrieval results or missing relevant content may drastically decrease the final performance, as the information provided to the generator may contain irrelevant objects or misleading information. \citet{Barnett2024Seven} also state that the reason for that might be the missing content, e.g., ``when asking a question that cannot be answered from the available documents''. The next failure point is called ``not in context'' \citet{Barnett2024Seven}. In this case, the extracted documents were not correctly consolidated during the post-retrieval process. The last three failure points relate to the generated output: the incorrect format of the output, incorrect specificity (``not specific enough or is too specific to address the user’s need''), and incomplete output that misses essential information even though being extracted by the retriever.  

When combining RAG with another system, the most common limitation is extra overhead: additional retrieval and interaction processes lead to increased latency of the system. Moreover, speed time also depends on the gap between retrievers and generators: the integration process and increased system complexity might be other bottlenecks that should be considered. When applying RAG to LLMs or other generators with a limited context size, lengthy context might become a problem: the models might not be able to accept the whole retrieved data as input and the generation process will take much more time than expected. 

\paragraph{$\bf D_{16}$ Future Directions:} The last dimension outlines future directions for the RAG systems based on the findings of \citet{Zhao2024Retrieval-Augmented}. The most straightforward directions are further development of RAG methodologies, enhancements, and applications. This might include new interactions between the retriever and generator, various modular RAG architectures with looping, and more advanced pre- and post-processing steps. Another direction is efficient deployment and processing. When discussing limitations, most of the integration limitations were related to efficiency and latency. Hopefully, future research on RAG capacities will allow shorter system response time and easier deployment. Another important research direction is the incorporation of long-tail and real-time knowledge. With the rapid growth of the data, it is extremely difficult to constantly update large retrievals in RAG. Many existing works apply a static database for knowledge retrieval, which requires re-indexing and/or computing additional representations. \citet{Zhao2024Retrieval-Augmented} expect newer techniques to solve the issue, as well as provide better retrieval of less commonly referenced data. Lastly, the combination of other techniques might be also seen as a promising direction, e.g. integration of RAG with the new state space model architecture like Mamba \cite{Gu2023Mamba:} or RWKV \cite{Peng2023RWKV:}. 

\section{Discussion} \label{sec:discussion}

RAG application systems are an emerging technology that has received considerable attention outside the NLP community, which addresses the limitations of LLM applications \cite{Gao2024Retrieval-Augmented,Leiser2024HILL:}. While recent studies address both the applications of LLMs and methods to mitigate their shortcomings,
RAGs have not yet been fully recognized outside NLP community or explored as a potential solution to these limitations. Thus, our taxonomy provides a basis for applying RAGs as an emerging technology for novel fields of application. Based on our taxonomy, we see that the broader community can engage in a socio-technical perspective for guiding future RAG applications. 


\paragraph{Domain-specific applications:} Our taxonomy shows that different mediums of generation are gaining interest, including conceptual modeling approaches \cite{Baumann2024Combining}. For example, process modeling already leverages generative AI for improving and (re-) designing organizational processes \cite{Dun2023ProcessGAN:}. RAGs appear to make such application systems much more viable, as our taxonomy provides us an example, where knowledge structure already incorporates conceptual process modeling types \cite{Baumann2024Combining}. Thus, we see potential in incorporating RAGs into design science research endeavors \cite{Hevner2004Design,Peffers2007Design,Teixeira2019Advancing} to address a variety of domain-specific applications. In our analyses, we identified proof-of-concepts and have seen little research on practical RAG applications. Thus, we call for future research to address this lack of proof-of-value \cite{Nunamaker2015last}. 

\paragraph{Business Value of RAG Applications:} The impacts of AI have largely been due to increasing business value following business processes \cite{Davenport2018Artificial}. Research into conversational agents has shown that they can lead to tangible business value \cite{Kull2021How,Mariani2023Artificial,MCLEAN2021312}, whereas the assessment of LLM-based impact of business value remains under-studied \cite{Storey2024design}. This could be attributed to current restraints of LLMs. However, with RAGs, there might be legitimate potential to create tangible business value, be it by also addressing knowledge- and labor-intensive services or improving work conditions.  

\paragraph{Ethical, legal, social implications:} Most current research is moving towards sustainability, including calls for studying social value \cite{Nunamaker2015last} and putting the need for reflecting on values in system designs (e.g.: \citet{Bednar2023Power,Friedman2013Value}). With LLMs already being discussed with ecological inefficiencies due to their potential high carbon emissions, integrating RAGs to improve LLM applications might have unforeseen consequences. Similarly, we see ongoing discussions on the societal implications of improving works systems, leading to potential job losses (e.g. \citet{Brynjolfsson2023Generative}) or legal disputes about leveraging copyrighted content to generate new content \cite{Golatkar2024CPR:,Samuelson2023Generative}. Integrating RAGs can improve either, yet its ethical, legal, and social implications require careful consideration, exemplifying its socio-technical nature. 

\paragraph{Digital transformation and RAG implementations:} The challenges of adopting technologies, including AI \cite{Grønsund2020Augmenting}, firms are facing include high resistance to change and organizational barriers \cite{Vial2019Understanding}. Since RAG applications address traditionally knowledge-intensive tasks, such as analysis and interpretation of data, aiming to outperform human capabilities, we see the potential that RAG applications can lead to increased organizational resistance, especially when integrated into existing Enterprise applications. RAGs can play a considerable role as orchestrators of enterprise application systems \cite{Böhmann2014Service} to call each functionality as part of its retrieval and generate respective outputs. This increasing complexity of heterogeneous applications might lead to new challenges for digital transformation.

Thus, the taxonomy is broad with sixteen dimensions. As the field is still evolving, we deem this initial breadth beneficial to shape our community understanding. While the field is settling and maturing, a narrowed-down taxonomy could be the next step, to further increase the conciseness and applicability, especially for practice. As of now, expert knowledge is still needed to assess several details within the taxonomy. 

\section{Conclusion} \label{sec:conclusion}
Our RAG taxonomy provides a structured way to categorize and analyze the diverse approaches, system features, and technologies that constitute RAG applications. Thus, we contribute to a clearer understanding of its components and their interactions. The taxonomy has five meta-dimensions, sixteen dimensions, and sixty-one characteristics, reflecting the inherent complexities of current RAGs. This systematic classification is essential for different researchers and practitioners to identify gaps in the current technology, facilitate research and development efforts, and identify potential use cases for real-world applications. Based on our taxonomy, we also present several avenues for future research, accommodating the RAG characteristics for different application types. Overall, the taxonomy not only enriches the academic discourse by providing a foundational framework for study and discussion but also guides practical implementations and innovations within the field. 


\section{Limitations}

Our taxonomy development approach has several limitations, which can be attributed to the novelty of our phenomenon of interest. 
Additionally, while dealing with pre-prints in a fast-moving research field, papers get updated while working with them, leading to inconsistencies for the research team, that need to be reworked afterward. Furthermore, we do not claim completeness, as the field is quickly moving forward, and we aim to capture an initial view of the emerging phenomenon, calling for future taxonomy extensions.

\bibliography{ref-extracts}

\clearpage
\onecolumn
\appendix

\section{Appendix}
\label{sec:appendix}

\begin{figure*}[ht!]
\centering
\includegraphics[trim={3cm 6cm 3cm 2cm}, width=0.9\textwidth]{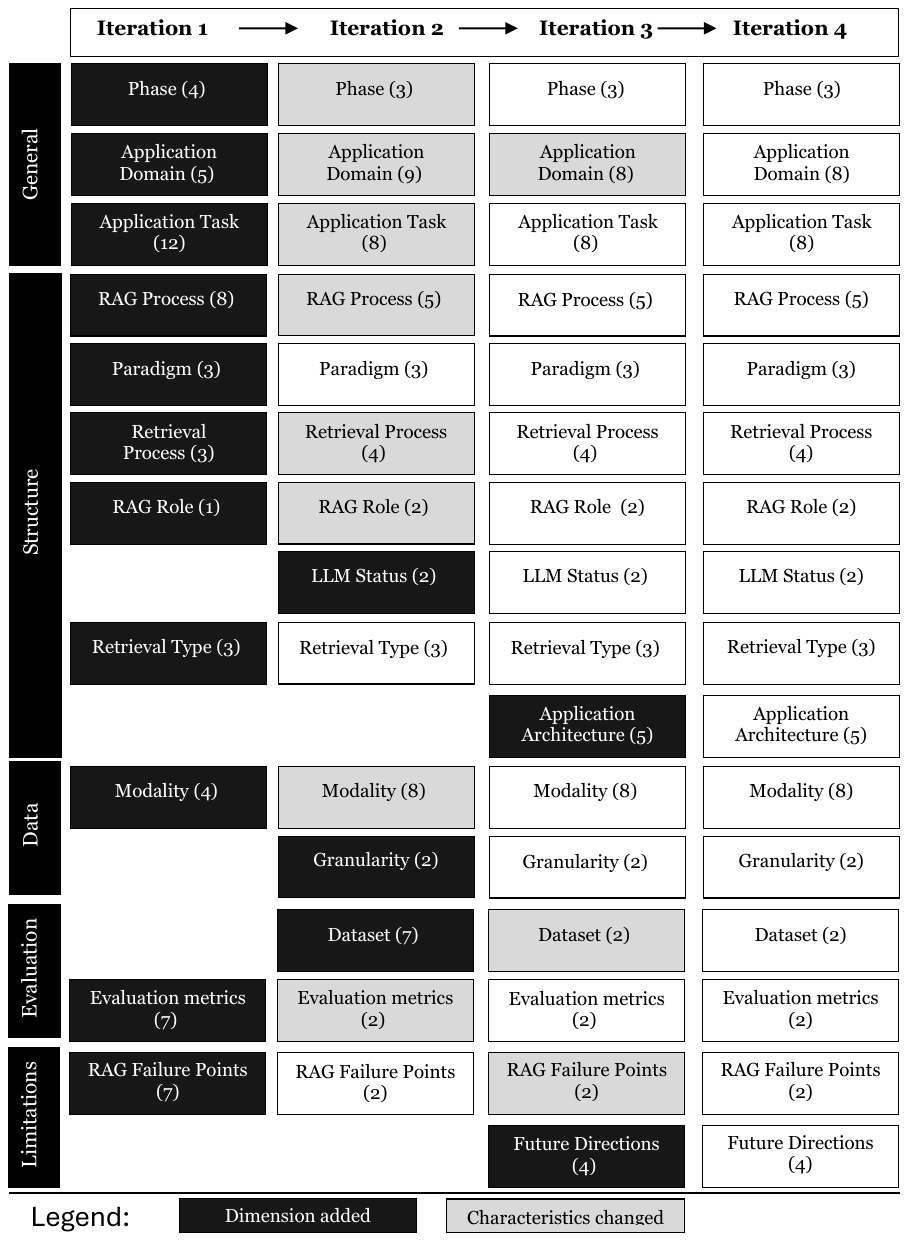}
\caption{Development of taxonomy dimensions and characteristics (adapted from \citet{Bräker2022Conceptualizing,Remane2016Taxonomy})}
    \label{fig:iterations}
\end{figure*}

\end{document}